\documentclass[11pt]{article}
\usepackage{amsmath} % for typesetting matrices
\usepackage{amssymb}
\usepackage{verbatim} % comment out chunks of file
\usepackage{bm} % bold math
\usepackage[colorlinks=true,citecolor=blue,linkcolor=black]{hyperref}
\usepackage{authblk} % several authors
\usepackage{sectsty} % change font size in section titles
\sectionfont{\large}
\subsectionfont{\normalsize}
\subsubsectionfont{\normalsize}
\numberwithin{equation}{section} % number equations by section

\def\Eq#1{Eq.~(\ref{#1})}
\def\aux{\tilde{\phi}} % auxiliary hydro fields
\def\k{{\bf k}}

\def\L{{\cal L}}
\def\vs{v_{\rm s}}
\def\coeff#1#2{{\textstyle {\frac {#1}{#2}}}}

\title{ Towards an effective action for relativistic dissipative hydrodynamics
      }

\author{Pavel Kovtun}
\affil{
    Department of Physics and Astronomy, University of Victoria, Victoria, BC, V8P 5C2, Canada}
\author{Guy D.\ Moore}
\affil{
    Department of Physics, McGill University, 3600 rue University, Montr\'{e}al, QC H3A 2T8, Canada}
\author{Paul Romatschke}
\affil{
    Department of Physics, 390 UCB, University of Colorado, Boulder, CO 80309-0390, USA\vspace{-4ex}}

\date{ (May 2014)}

\begin{document}

\maketitle

\begin{abstract}
\noindent
We propose an effective action for first order relativistic dissipative
hydrodynamics that can be used to evaluate $n$-point symmetrized
correlation functions, taking into account thermal fluctuations of the
hydrodynamic variables.
\end{abstract}

\section{Introduction}

The study of fluid dynamics is a centuries-old discipline that still
seems nowhere near closure. The  ``classical'' relativistic fluid
dynamics can be derived by requiring conservation of the energy-momentum
tensor and global symmetry currents \cite{LL6}. In modern understanding,
these conservation equations should be constructed order by order in the
derivative expansion of the hydrodynamic variables, similar to the
derivative expansion in effective field theory.  Truncating the
derivative expansion at first order, second order and higher order one
obtains ideal fluid dynamics (Euler equations), viscous fluid dynamics
(Navier-Stokes equations), and higher-order dissipative
hydrodynamics~\cite{Bhattacharyya:2008jc,Baier:2007ix}. The classical
hydrodynamic correlation functions can then be obtained by varying with
respect to external sources, see e.g.~\cite{Kovtun:2012rj}.

In non-relativistic fluids, it is well known that there are correlation
functions of the hydrodynamic variables which can not be reproduced by
such classical hydrodynamic equations, even when the frequency and
momentum are arbitrarily small~\cite{Pomeau197563}. The reason is that
while the classical equations describe flows generated by external
sources, they neglect hydrodynamic excitations generated by thermal
fluctuations within the fluid. These effects may be taken into account
by supplementing the classical hydrodynamic equations with stochastic
noise terms whose correlation functions are taken to be Gaussian white
noise~\cite{Pomeau197563}. It is natural to expect that a similar
stochastic modification is required for relativistic fluids in order to
correctly reproduce physical observables.

In linear non-relativistic hydrodynamics, such noise terms were
introduced long ago by Landau and Lifshitz~\cite{LL-1957}. When
considering the full non-linear theory of stochastic hydrodynamics, one
finds that the interactions lead to changes in the basic parameters of
the classical theory, such as the shear viscosity
coefficient~\cite{Forster:1977zz}. More generally, correlation functions
evaluated in stochastic hydrodynamics will differ from their classical
counterparts by fluctuation corrections involving loops of the
hydrodynamic modes. Stochastic equations for hydrodynamic variables can
be readily converted to a functional integral
form~\cite{PhysRevB.18.353}, providing one with an effective field
theory.  The purpose of this note is to write down an effective action
for dissipative relativistic fluids.

We emphasize that our interest is not in an action that will give rise
to the classical hydrodynamic equations upon using a variational
procedure. Rather, we are interested in an action which can be used in a
standard way in the functional integral to evaluate hydrodynamic
correlation functions. While it is straightforward to derive such an
effective action for the linearized viscous relativistic
hydrodynamics~\cite{Kovtun:2012rj}, the full non-linear hydrodynamics
and the derivative expansion require more work. The fields in the
effective theory include the hydrodynamic variables (fluid velocity,
temperature etc), and we will refer to this effective theory as
``statistical hydrodynamics'', to distinguish it from classical
hydrodynamics which ignores fluctuations. The 1PI effective action of
statistical hydrodynamics should give rise to the classical hydrodynamic
equations at tree level, but will contain corrections to classical
hydrodynamics once the loops are taken into account. The loops here are
not the quantum loops (as one is not quantizing the classical
hydrodynamics), but rather reflect statistical fluctuations of the
hydrodynamic variables.

We pause to comment on previous work addressing related questions.
A variational formulation of classical ideal relativistic hydrodynamics
(neglecting the derivative expansion, fluctuations, and dissipation) is
an old subject discussed by many authors in various forms, see
e.g.~\cite{Schutz:1970my, Brown:1992kc, Carter:1994rv}. As mentioned
above, we don't expect such classical constructions to be helpful for
statistical hydrodynamics.
Refs.~\cite{Dubovsky:2011sj, Bhattacharya:2012zx} studied effective
actions for relativistic fluids, taking into account the derivative
expansion, however the resulting effective action only captured
non-dissipative information. Similarly, Refs.~\cite{Banerjee:2012iz,
  Jensen:2012jh} derived generating functionals of relativistic fluids
coupled to external sources in equilibrium. Again, this allowed a
systematic construction to any order in the derivative expansion, but
only captured static non-dissipative physics. For variational approaches
aiming to incorporate dissipation in classical hydrodynamics, see
e.g.~\cite{Andersson:2006nr, Endlich:2012vt, Grozdanov:2013dba}.

Recently, there have also been efforts to understand dissipation in
relativistic statistical hydrodynamics (with fluctuation corrections),
partly motivated by the experimental study of the quark-gluon plasma in
heavy-ion collisions. Refs.~\cite{Kovtun:2003vj, PeraltaRamos:2011es,
  Kovtun:2011np, Kovtun:2012rj} looked at statistical one-loop
corrections to the shear viscosity, but lacked a systematic
field-theoretic framework. See~\cite{Kapusta:2011gt, Kumar:2013twa,
  Murase:2013tma,Young:2013fka} for other recent work on relativistic
fluctuating
hydrodynamics, including the Israel-Stewart formulation. It is worth
pointing out that the fluctuation corrections render the derivative
expansion in purely classical relativistic hydrodynamics
ill-defined~\cite{Kovtun:2011np}. Clearly, one needs a unified
calculational framework that takes into account the full non-linearity
of relativistic hydrodynamics, the derivative expansion, and
fluctuations of the hydrodynamic variables. The present paper is a step
in this direction.

\section{Noisy hydrodynamics}
\subsection{Setup}
\label{sec:setup}
Classical relativistic hydrodynamics~\cite{LL6} is a set of partial
differential equations for the hydrodynamic fields $u^\mu(x)$, $T(x)$,
and (for fluids with a global $U(1)$ charge) $\mu(x)$. Collectively
denoting these hydrodynamic fields as $\phi$, we will write the
classical hydrodynamic equations in the form $E^a(\phi)=0$, where $E^\mu
= \partial_\nu T^{\nu\mu}_{\rm cl}$, $E^{d+1}=u^2+1$, $E^{d+2} =
\partial_\mu J^\mu_{\rm cl}$, and $d$ is the number of spatial
dimensions. Here $T^{\mu\nu}_{\rm cl}$ and $J^\mu_{\rm cl}$ are the
(symmetric) energy-momentum tensor and the $U(1)$ global symmetry
current, given in terms of $\phi$. The fluid velocity is normalized%
\footnote{%
  Our metric signature is $[{-}{+}{+}{+}]$, \textsl{eg}, space-positive.}
as $u^2=-1$, $T$ is the temperature, and $\mu$ is the chemical
potential. The constitutive relations expressing $T^{\mu\nu}_{\rm cl}$
and $J^\mu_{\rm cl}$ in terms of $\phi$ are normally written in a given
``frame'' (a particular out-of equilibrium definition of $\phi$), to a
given order in the derivatives of~$\phi$. The starting point for the
stochastic hydrodynamics is the modification of the classical
hydrodynamic equations by ``noise'' terms which are interpreted as
microscopic stresses and currents~\cite{LL-1957}, so that the
hydrodynamic equations take the form
$\partial_\mu T^{\mu\nu}=0$, and $\partial_\mu J^\mu = 0$, where
$T^{\mu\nu} = T^{\mu\nu}_{\rm cl} + \tau^{\mu\nu}$, and $J^\mu =
J^\mu_{\rm cl}+r^\mu$. The microscopic contributions
$\tau^{\mu\nu}(\phi,\xi)$ and $r^\mu(\phi,\xi)$ are functionals of both
the hydrodynamic fields $\phi$ and the noise fields collectively denoted
as $\xi$, so that the hydrodynamic equations become stochastic equations

\begin{equation}
  E^a(\phi) + f^a(\phi,\xi) = 0\,,
\label{eq:EOM-noise}
\end{equation}
where $f^\mu = \partial_\nu \tau^{\nu\mu}$ and $f^{d+2} = \partial_\mu r^\mu$.
The form of the force $f^a$ and the dynamics of the noise fields need to
be determined by the problem at hand. In particular, they must be such
that the fluctuation-dissipation theorem is satisfied in equilibrium.

One can convert \Eq{eq:EOM-noise} to a functional integral
form. Let us denote the solution to \Eq{eq:EOM-noise} as
$\phi_\xi$. Upon solving \Eq{eq:EOM-noise}, the energy-momentum
tensor and the current will become functionals of the noise,
$T^{\mu\nu}[\phi_\xi,\xi]$, $J^\mu[\phi_\xi,\xi]$.  For a general
function $O(\phi_\xi)$ we have
$$
  O(\phi_\xi) = \int \!\!D\phi\; \delta\!\left(E^a(\phi) + f^a(\phi,\xi)
  \right) J(\phi,\xi) O(\phi)\,,
$$
where the Jacobian is $J={\rm det}
\frac{\delta(E^a+f^a)}{\delta\phi_b}$. If the dynamics of $\xi$ is
independent of $\phi$, so that the noise average is performed with some
$\phi$-independent action $S_{\rm n}[\xi]$, the correlation functions
can be written as
\begin{equation}
  \langle T^{\mu\nu} T^{\alpha\beta} \dots \rangle =
  \int\! D\xi\, D\phi \, D\aux\;
  e^{\,i\int \aux_a [E^a(\phi) + f^a(\phi,\xi)]}\;
  J(\phi,\xi)\;
  e^{-S_{\rm n}[\xi]}\;
   T^{\mu\nu}[\phi,\xi]\, T^{\alpha\beta}[\phi,\xi] \dots .
\label{eq:TT-1}
\end{equation}
The corresponding partition function is
\begin{equation}
  Z =
  \int\! D\xi\, D\phi \, D\aux\;
  e^{\,i\int \aux_a [E^a(\phi) + f^a(\phi,\xi)]}\;
  J(\phi,\xi)\;
  e^{-S_{\rm n}[\xi]}\,.
\label{eq:Z0}
\end{equation}
Alternatively, one can define stochastic hydrodynamics by the functional
integral representation,
\begin{equation}
  Z = \int\! D\xi\, D\phi \, D\aux\;
  e^{\,i\int \aux_a [E^a(\phi) + f^a(\phi,\xi)]}\;
  e^{-S_\xi[\phi,\xi]}\,,
\label{eq:Z-hydro}
\end{equation}
where the auxiliary fields $\aux_a$ ensure that \Eq{eq:EOM-noise} is
satisfied, and the noise action $S_\xi$ needs to be specified. Normally,
the central limit theorem is invoked to argue that the noise is
Gaussian, hence the noise action is quadratic in $\xi$. In this case
$\xi$ can be integrated out, leaving one with the effective action
$S_{\rm eff}(\phi,\aux)$. A proposal for the effective action in
stochastic hydrodynamics amounts to a choice of $f^a$ and $S_\xi$.

The functional integral \Eq{eq:Z-hydro} can in principle be used to
compute correlation functions of the hydrodynamic fields, and hence of
$T^{\mu\nu}$ and $J^\mu$. As the order of the fields does not matter
inside the functional integral, these are unordered (or symmetrized)
correlation functions.%
\footnote{
	 The effective theory discussed here is supposed to be valid in
         the hydrodynamic limit $\omega\to0$. In equilibrium, the
         difference between unordered and symmetrized functions is
         $O(\omega/T)$ for $\omega\ll T$. Out of equilibrium, we assume
         that there is a scale $\omega_0$ such that the difference
         between unordered and symmetrized functions is negligible for
         $\omega\ll\omega_0$.
	 }
The effective action given by \Eq{eq:Z-hydro} contains extra fields,
in addition to the hydrodynamic fields $\phi$. The extra fields can be
thought of as ``degrees of freedom'' giving rise to dissipation. As the
effective theory \Eq{eq:Z-hydro} describes dissipative physics, the
effective action need not be real.

In what follows we will apply the formulation \Eq{eq:Z-hydro} to the
first-order hydrodynamics in the Landau-Lifshitz
``frame''~\cite{LL6}. The classical constitutive relations can be taken
as
\begin{subequations}
\label{eq:const-rel}
\begin{align}
  & T^{\mu\nu}_{\rm cl} = \epsilon u^\mu u^\nu + p \Delta^{\mu\nu}
     - G^{\mu\nu\alpha\beta} \partial_\alpha u_\beta \,,\\[5pt]
  & J^\mu_{\rm cl} = n u^\mu
     -\sigma T \Delta^{\mu\nu} \partial_\nu (\mu/T)
     \,,
\end{align}
\end{subequations}
with $\epsilon$, $p$, and $n$ the equilibrium energy density, pressure,
and charge density, and with the last terms describing the dissipative
part of the dynamics.
Here $\Delta^{\mu\nu} \equiv \eta^{\mu\nu} + u^\mu u^\nu$ is the
projector to the space components of the local rest frame, and
$G^{\mu\nu\alpha\beta}\equiv 2\eta S_{\rm T}^{\mu\nu\alpha\beta} 
 +d \zeta S_{\rm L}^{\mu\nu\alpha\beta}$, where
$S_{\rm T}^{\mu\nu\alpha\beta} \equiv \frac12(\Delta^{\mu\alpha}
\Delta^{\nu\beta} + \Delta^{\mu\alpha} \Delta^{\nu\beta} - \frac2d
\Delta^{\mu\nu} \Delta^{\alpha\beta})$ and
$S_{\rm L}^{\mu\nu\alpha\beta} 
   \equiv \frac{1}{d} \Delta^{\mu\nu} \Delta^{\alpha\beta}$
are transverse and longitudinal spatial projectors.
$\eta(T,\mu)$ is the shear viscosity,
$\zeta(T,\mu)$ is the bulk viscosity, and $\sigma(T,\mu)$ is the charge
conductivity. Working in the Landau-Lifshitz frame, we will impose
$u_\mu \tau^{\mu\nu}=0$ and $u_\mu r^\mu=0$.

\subsection{Linear fluctuations in equilibrium}
To illustrate the general procedure, let us look at small fluctuations
in thermal equilibrium with constant $\bar T$,  constant $\bar\mu=0$,
and constant $\bar u^\mu=(1,{\bf 0})$. To linear order in fluctuations
in the Landau-Lifshitz frame $\tau^{0\mu}=0$, $r^0=0$, and the
constitutive relations become
\begin{align*}
  & T_{ij}
    = \delta_{ij} (\bar p + \bar s\, \delta T)
    - \bar\eta(\partial_i v_j + \partial_j v_i - \coeff2d \delta_{ij}
    \partial_k v^k)
    - \bar\zeta \delta_{ij} \partial_k v^k
    +\tau_{ij}\,,\\
  & J_i = -\bar\sigma \partial_i \mu + r_i\,.
\end{align*}
To linear order in fluctuations, $\tau_{ij}$ and $r_i$ do not depend on
the hydrodynamic fields and can be treated as external sources. For the
Fourier components it is then straightforward to find
\begin{align*}
  &  \delta T(\omega,\k)
     = \frac{1}{\partial\bar\epsilon/\partial \bar T}
     \frac{k_i k_j \tau^{ij}}{\omega^2 - \vs^2 \k^2 + i\gamma_s \omega\k^2}\,,\\[5pt]
  &  v^i(\omega,\k) = \left( \delta^{ij} - \frac{k_i k_j}{\k^2} \right)
     \frac{k_m \tau^{jm}}{\bar w (\omega + i\gamma_\eta \k^2)}
     +\frac{k^i}{\k^2} \frac{\omega}{\bar w}
     \frac{k_m k_n \tau^{mn}}{\omega^2 - \vs^2 \k^2 + i\gamma_s \omega\k^2}\,,\\[5pt]
  &  \mu(\omega,\k) = \frac{k_m r^m}{\bar\chi(\omega + iD\k^2)}\,,
\end{align*}
where $\vs^2 = s/(\partial\bar\epsilon/\partial \bar T)$ is the speed of
sound squared, $\gamma_\eta\equiv \bar\eta/\bar w$, $\gamma_\zeta\equiv
\bar\zeta/\bar w$, $\gamma_s\equiv \gamma_\zeta+
\frac{2d-2}{d}\gamma_\eta$, $\bar w\equiv \bar\epsilon+\bar p$ is the
equilibrium enthalpy density, $\bar\chi\equiv (\partial\bar
n/\partial\mu)_{\mu=0}$ is the equilibrium charge susceptibility, and
$D=\bar\sigma/\bar \chi$ is the charge diffusion constant. The
symmetrized two-point functions can be evaluated provided one specifies
the dynamics of $\tau^{ij}$ and $r^i$. In our case they are taken as
Gaussian fields with~\cite{LL9}
\begin{subequations}
\label{eq:linear-noise}
\begin{align}
  & \langle r_i (x) r_j(y) \rangle = 2\bar T \bar\sigma \delta_{ij} \delta(x-y)\,,\\[5pt]
  & \langle \tau_{ij}(x) \tau_{kl}(y) \rangle = 2\bar T
    G_{ijkl}
    \delta(x-y)\,,
\end{align}
\end{subequations}
where $G_{ijkl} =  2\bar\eta S_{{\rm T}\,ijkl} 
+ d \bar\zeta S_{{\rm L}\,ijkl}$ as above.  Choosing $\k$ along ${\bf
  z}$ gives the usual Kubo formula in terms of the symmetrized function
of $T_{xy}$,
$$
  G_{xy,xy}^{S}(\omega,\k) = 2\bar T \bar\eta \,.
$$
The above noise average can be represented with a Gaussian functional integral,
$$
  \langle \dots \rangle = \int\! D\tau_{ij}
  \!\!\!\!\!\!\!\!\!\!\!\raisebox{-2.5ex}{ ${\scriptstyle i\leqslant j}$ }\,
  Dr_k\;
  e^{-S_\xi[r,\tau]} \dots
$$
where
$$
  S_\xi[r,\tau] = \frac12 \int\!\! dt\, d^dx \left(
  \frac{r_i \bar\sigma^{-1} r_i}{2\bar T}
  + \frac{\tau_{ij} G^{-1}_{ijkl} \tau_{kl}}{2\bar T} \right)\,,
$$
and
$G^{-1}_{ijkl} = \frac{1}{2 \bar\eta} S_{{\rm T}\,ijkl} 
+ \frac{1}{ d \bar\zeta}S_{{\rm L}\,ijkl}$.
The action $S_\xi$ is positive definite (except for the trivial
configuration $r_i=0$, $\tau_{ij}=0$), and in the case of linear
fluctuations does not depend on the hydrodynamic fields $\phi$.
Integrating out the noise fields $r_i$ and $\tau_{ij}$ in
\Eq{eq:Z-hydro} gives
$$
  Z = \int\!D\phi\,D\aux\; e^{-S_{\rm eff}[\phi,\aux]}\,,
$$
where
\begin{equation}
  S_{\rm eff}[\phi,\aux] = \int\!dt\,d^dx \left[
    i\,   \partial_\mu \aux_\nu\,  T^{\mu\nu}_{\rm cl}\,
  + \bar T \partial_i \aux_j G_{ijkl}\, \partial_k \aux_l
  + i\,   \partial_\mu \aux_{d+2} \,J^\mu_{\rm cl}\,
  +\bar T \sigma  \delta_{ij} \partial_i \aux_{d+2}\,\partial_j \aux_{d+2}
  \right]\,
\label{eq:Seff-linear}
\end{equation}
is the effective action for linear viscous hydrodynamics~\cite{Kovtun:2012rj}.
The hydrodynamic fields are $\phi=(v_i,\delta T,\mu)$, and the stress
tensor and the current are given by the classical linear constitutive
relations, e.g. $T^{ij}_{\rm cl} = \delta_{ij} (\bar p + \bar s\, \delta
T) - G^{ijkl} \partial_k v_l$. The equilibrium correlation functions of
$T^{\mu\nu}$ and $J^\mu$ are straightforwardly evaluated using the
effective action~\Eq{eq:Seff-linear}. As expected, the effective
action is local, but not real. Integrating out auxiliary fields $\aux_a$
will give rise to an action which is real, but non-local. If the
auxiliary fields are rescaled as $\aux\to \bar T \aux$, the action can
be written as $S_{\rm eff} = (1/{\bar T}) \int\dots$, signifying that
$\bar T$ determines the strength of thermal fluctuations.

\subsection{The covariant form}
Beyond the linear approximation, we use $\phi=(u^\lambda,T,\mu)$ as the
hydrodynamic fields. The quadratic effective
action~\Eq{eq:Seff-linear} na\"ively generalizes to
\begin{equation}
  S_{\rm eff}[\phi,\aux] = \int\!dt\,d^dx \left[
    i\,   \partial_\mu \aux_\nu\,  T^{\mu\nu}_{\rm cl}\,
  + T \partial_\mu \aux_\nu G^{\mu\nu\alpha\beta}\, \partial_\alpha \aux_\beta
  + i\,   \partial_\mu \aux_5 \,J^\mu_{\rm cl}\,
  + T \sigma   \Delta^{\mu\nu} \partial_\mu \aux_5  \partial_\nu \aux_5
  \right] .
\label{eq:Seff-2}
\end{equation}
Here $T^{\mu\nu}_{\rm cl}$ and $J^\mu_{\rm cl}$ are given by
\Eq{eq:const-rel}, $\aux_5\equiv \aux_{d+2}$, and the constraint
$u^2=-1$ is implied. The effective action is not real, nor should it
be. The action is invariant under complex conjugation combined with
$\aux_\mu \to -\aux_\mu$, $\aux_5 \to -\aux_5$.

The auxiliary fields $\aux_\mu$ and $\aux_5$ are derivatively coupled,
suggesting a Goldstone-boson interpretation, similar
to~Ref.~\cite{Endlich:2012vt}. The Noether currents of the shift
symmetry are
\begin{subequations}
\label{eq:TJNoether-1}
\begin{align}
  & T^{\mu\nu} = T^{\mu\nu}_{\rm cl} -2iT G^{\mu\nu\alpha\beta}
  \partial_\alpha \aux_\beta\,,\\[5pt]
  & J^\mu = J^\mu_{\rm cl} -2iT \sigma \Delta^{\mu\nu} \partial_\nu \aux_5\,\,,
\end{align}
\end{subequations}
and correspond to the full energy-momentum tensor and the current.

While this development is suggestive, we have not derived it, and in
fact there are some problems, which we now enumerate:
\begin{itemize}
\item
The relativistic version of \Eq{eq:linear-noise} is
\begin{subequations}
\label{eq:noise-2}
\begin{align}
  & \langle r_\mu (x) r_\nu(y) \rangle = 2  T \sigma \Delta_{\mu\nu} \delta(x-y)\,,\\[5pt]
  & \langle \tau_{\mu\nu}(x) \tau_{\alpha\beta}(y) \rangle = 2 T
    G_{\mu\nu\alpha\beta}
    \delta(x-y)\,,
\end{align}
\end{subequations}
indicating that the noise is not independent of $\phi$, contrary to the
assumptions made before \Eq{eq:TT-1}.  This can be
fixed by rescaling the noise fields $r_\mu$ and $\tau_{\mu\nu}$ by the
``square root'' of the coefficients appearing in the right-hand side
of~\Eq{eq:noise-2}. The conservation equations
$\partial_\mu(T^{\mu\nu}_{\rm cl} + \tau^{\mu\nu})=0$ and
$\partial_\mu(J^\mu_{\rm cl}+r^\mu)=0$ become stochastic differential
equations with multiplicative noise.  Ambiguities in defining such
stochastic differential equations must then be resolved in establishing
the form of the functional integral.

\item
If an effective action is to be derived from a stochastic differential
equation, there has to be the corresponding Jacobian, as indicated in
\Eq{eq:Z0}. In linear hydrodynamics, the Jacobian can be dropped
because it is field-independent, which is not the case in non-linear
hydrodynamics. Such a Jacobian is ignored in~\Eq{eq:Seff-2}.
\end{itemize}
We now turn to the above points.

\section{The effective action}

\subsection{The noise}
In order to arrive at the effective action, one can start from classical
hydrodynamics augmented with noise terms. Rather than using the
correlations \Eq{eq:noise-2}, one has to redefine the noise so that
the noise action does not depend on the hydrodynamic variables. To do
so, we introduce a noise field $\xi_\mu$ with $\langle \xi_\alpha(x)
\xi_\beta(x')\rangle = \eta_{\alpha\beta} \delta(x-x')$, and a symmetric
noise field $\xi_{\mu\nu}$ with $\langle \xi_{\mu\nu}(x)
\xi_{\alpha\beta}(x')\rangle = \frac12(\eta_{\mu\alpha} \eta_{\nu\beta}
+ \eta_{\nu\alpha}\eta_{\mu\beta}) \delta(x-x')$. The corresponding
contributions to the energy-momentum tensor and the current are
\begin{align}
  & \tau_{\mu\nu} = \sqrt{2T} G^{1/2}_{\mu\nu\alpha\beta}\,
  \xi^{\alpha\beta}\,,\ \ \ \
  r_\mu = \sqrt{2T\sigma}\, \Delta_{\mu\nu}\,\xi^\nu \,,
\end{align}
where $G^{1/2}_{\mu\nu\alpha\beta} = \sqrt{2\eta}\,
S_{{\rm T}\,\mu\nu\alpha\beta} 
  + \sqrt{d\zeta}\, S_{{\rm L}\,\mu\nu\alpha\beta}$
satisfies $G^{1/2}_{\mu\nu\alpha\beta} G^{1/2\,
  \alpha\beta\rho\sigma} = G_{\mu\nu}^{\ \ \ \rho\sigma}$, and
$G_{\mu\nu\rho\sigma}$ is defined below \Eq{eq:const-rel}. The
construction of the effective action can now proceed as described in
Sec.~\ref{sec:setup}, with%
\footnote{
        Note that $\xi_\mu$ and $\xi_{\mu\nu}$ contain negative-norm
        components.  We implicitly assume that this is handled through
        continuation of some noise components to imaginary values,
        so the functional weight bounds the integrals.  The projection
        operators ensure that wrong-norm components are never
        physically relevant.
}
$$
  S_{\rm n}[\xi] = \frac12 \int\!\! dt\,d^d x
  \left(\xi_\mu \xi^\mu + \xi_{\mu\nu} \xi^{\mu\nu} \right)\,.
$$
Ignoring the Jacobian in \Eq{eq:Z0} and integrating out $\xi_\mu$ and
$\xi_{\mu\nu}$ gives the effective action \Eq{eq:Seff-2}.

\subsection{The discretization}

\Eq{eq:EOM-noise} as written is ambiguous because Langevin equations
must generally be written as discrete-time equations, and the continuous
time limit can depend on the manner of the discretization.
This is particularly true for the case of multiplicative noise
(see for instance Ref.~\cite{Arnold:1999va}).  The essential feature of
hydrodynamic equations is that current and stress conservation must
be exact statements.  For a discretization with time spacing
$\Delta_t$ and spatial spacing $\Delta_x$, and with $\phi$ defined on
sites, we believe this should be achieved by defining
$J^\mu$ and $T^{\mu\nu}$ on the $\mu$-link, so that
\begin{equation}
\partial_\mu J^\mu(x) \equiv \sum_\mu \frac{J^\mu(x+\hat\mu\Delta_\mu/2) -
  J^\mu(x-\hat\mu \Delta_\mu/2)}{\Delta_\mu} \,,
\end{equation}
and similarly for $\partial_\mu T^{\mu\nu}$.

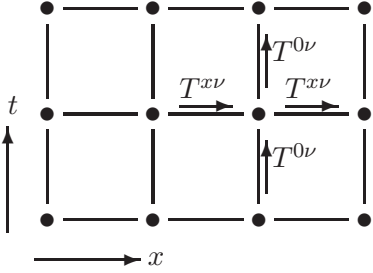
\begin{figure}[t]
\centerline{\begin{picture}(140,110)
  \thicklines
  \put(10,10){\vector(1,0){40}}
  \put(53,8){$x$}
  \put(0,20){\vector(0,1){40}}
  \put(0,65){$t$}
  \multiput(15,25)(40,0){4}{
    \multiput(0,0)(0,40){3}{\circle*{5}}}
  \multiput(21,25)(40,0){3}{
    \multiput(0,0)(0,40){3}{\line(1,0){28}}}
  \multiput(15,31)(40,0){4}{
    \multiput(0,0)(0,40){2}{\line(0,1){28}}}
  \multiput(100,45)(0,40){2}{$T^{0\nu}$}
  \multiput(98,35)(0,40){2}{\vector(0,1){20}}
  \multiput(65,71)(40,0){2}{$T^{x\nu}$}
  \multiput(65,68)(40,0){2}{\vector(1,0){20}}
\end{picture}}
\caption{\label{fig:discrete}
How $T^{\mu\nu}$ should be discretized.  $T^{\mu\nu}$ is the flux
of $P^\nu$ from one site to the neighboring site in the 
$\mu$-direction.  Stress-energy conservation at a site is the equality of
the sum of all incoming $P^\nu$ contributions and the sum of all
outgoing $P^\nu$ contributions.}
\end{figure}

Current conservation at a site is the vanishing of the signed sum of
currents onto and off of that site, which implements conservation
exactly. This is illustrated in Figure~\ref{fig:discrete}. The current on a link should be defined using the average of
$\phi$ at each end of the link, \textit{eg},%
\footnote{%
    One feature of this definition is that particle number and
    4-momentum density are not defined on sites, but on the temporal
    links between sites.  Charge conservation means that $J^0$ summed
    over temporal links at time $t-\Delta_t/2$ equals $J^0$ summed over
    temporal links at $t+\Delta_t/2$.}
$$
J^0(x+\hat t \Delta_t/2) = \left( \frac{n(x)+n(x+\hat t \Delta_t)}{2} \right)
          \left(\frac{u^0(x)+u^0(x+\hat t \Delta_t)}{2}\right) -{} \ldots \,.
$$
We will assume that this is the prescription for defining the discrete
equations of motion.  The subsequent steps we describe should then be
performed on these spacetime-discretized equations, with the continuum
limit taken at the end (if at all).  We expect this procedure to resolve
discretization ambiguities, and we implicitly assume that it has been
used in the following.

\subsection{The Jacobian}

We next turn to the incorporation of the Jacobian of \Eq{eq:Z0} into
the effective action.
The Jacobian is $J = \det J_{ab}(\phi,\xi)$, where
$J_{ab}=\delta(E^a{+}f^a)/\delta\phi_b$ is a
differential operator linear in $\xi$. The Jacobian can be exponentiated
by using ghost fields $\bar\psi_a$, $\psi_a$ as
$$
  J = \int\!\!D\bar\psi\,D\psi\; e^{-S_{\rm det}}\,,
$$
where $S_{\rm det} = \int\! dt\,d^dx\, \bar\psi_a J_{ab}\psi_b$. As
$J_{ab}$ is linear in $\xi$, this action can be written as
$$
  S_{\rm det} = \int\! dt\,d^dx\, \left(
  \xi_\mu F^\mu(\phi,\bar\psi,\psi) + \xi_{\alpha\beta}
  F^{\alpha\beta}(\phi,\bar\psi,\psi)
  + \bar\psi_a \frac{\delta E^a}{\delta\phi_b} \psi_b
  \right)\,,
$$
where $F^\mu(\phi,\bar\psi,\psi)$ and
$F^{\alpha\beta}(\phi,\bar\psi,\psi)$ are straightforward to evaluate,
for given a choice of the hydro variables $\phi_a$. Choosing $\phi =
(u,T,\mu)$, we have%
\begin{align*}
  & F^{\mu} = -\partial_\lambda \bar\psi_{d+2} \left[
    \frac{\delta C^{\lambda\mu}}{\delta u_\nu} \psi_\nu +
    \frac{\delta C^{\lambda\mu}}{\delta T} \psi_{d+1} +
    \frac{\delta C^{\lambda\mu}}{\delta \mu} \psi_{d+2}
    \right]\,,\\[5pt]
  & F^{\alpha\beta} = -\partial_\lambda \bar\psi_{\mu} \left[
    \frac{\delta C^{\lambda\mu\alpha\beta}}{\delta u_\nu} \psi_\nu +
    \frac{\delta C^{\lambda\mu\alpha\beta}}{\delta T} \psi_{d+1} +
    \frac{\delta C^{\lambda\mu\alpha\beta}}{\delta \mu} \psi_{d+2}
    \right]\,,
\end{align*}
with $C^{\mu\nu} \equiv \sqrt{2T\sigma}\Delta^{\mu\nu}$,
$C_{\mu\nu\alpha\beta} \equiv \sqrt{2T} G^{1/2}_{\mu\nu\alpha\beta}$.
Integrating out the noise $\xi_\mu$ and $\xi_{\mu\nu}$ gives the
effective Lagrangian
\begin{align}
  \L_{\rm eff} & =
  i\,   \partial_\mu \aux_\nu\,  T^{\mu\nu}_{\rm cl}\,
  - \coeff12 (F_{\alpha\beta} {+} i P_{\alpha\beta}) (F^{\alpha\beta}
  {+} iP^{\alpha\beta}) \nonumber\\
  & + i\,   \partial_\mu \aux_5 \,J^\mu_{\rm cl}\,
  - \coeff12 (F_\mu {+} i P_\mu) (F^\mu {+} iP^\mu)
  + \bar\psi_a \frac{\delta E^a}{\delta\phi_b} \psi_b \,,
\label{eq:Leff-1}
\end{align}
where $P^\mu \equiv \sqrt{2T\sigma}\,
\Delta^{\mu\lambda}\partial_\lambda\aux_5$, and $P^{\mu\nu} \equiv
\sqrt{2T}\, G^{1/2\, \mu\nu\lambda\sigma}
\partial_\lambda\aux_\sigma$.

\section{Conclusions}
Our proposal for the effective action in stochastic relativistic
hydrodynamics is
\begin{equation}
\label{eq:finalaction}
  S_{\rm eff}[\phi,\tilde\phi,\dots] = \!\int\!\! dt\, d^dx
  \left[ i\partial_\mu \tilde\phi_\nu\, T^{\mu\nu}_{\rm cl}
  + T G^{\mu\nu\lambda\sigma} \partial_\mu \tilde\phi_\nu\,
  \partial_\lambda \tilde\phi_\sigma
  + i\partial_\mu \tilde\phi_5\, J^\mu_{\rm cl}
  + T\sigma \Delta^{\mu\nu} \partial_\mu \tilde\phi_5\, \partial_\nu
  \tilde\phi_5
  + \dots
  \right]
\end{equation}
where the classical energy-momentum tensor and the current are given
by~\Eq{eq:const-rel}.  The hydrodynamic variables are
$\phi^a=(u^\lambda,T,\mu)$, and the $\tilde\phi$'s are the corresponding
auxiliary fields. The constraint $u^2=-1$ is implied. The dots
in~\Eq{eq:finalaction} denote ghost terms. The derivation
of~\Eq{eq:finalaction} parallels existing work in simpler
hydrodynamic systems. Several comments are in order.
\begin{itemize}
\item
Evaluating the 1PI effective action in a given background in the
theory~\Eq{eq:finalaction} should give rise to classical hydrodynamic
equations plus loop corrections. Among other things, the loop
corrections will renormalize the transport coefficients $\eta$, $\zeta$,
and $\sigma$, similar to what happens in simpler
systems~\cite{Forster:1977zz, Zanella:2002hh}.

\item
In the simplest case of a scale-invariant uncharged fluid in thermal
equilibrium, the effective action has only three parameters: equilibrium
temperature $T$, equilibrium entropy density $s$, and equilibrium shear
viscosity $\eta$. There is only one combination, $\lambda\equiv
(T/s^{1/d})\, (\eta/s)^{-1}$, which is dimensionless in the natural
units $c=1$ (the effective theory is classical, so there is
no~$\hbar$). In a large-$N$ gauge theory, $\lambda\to0$ as
$N\to\infty$. One expects that fluctuation corrections will be
suppressed by a positive power of $\lambda$, as happens in simpler
models.

\item
We have only taken into account first-order gradient terms in the
hydrodynamic equations of motion~\Eq{eq:const-rel}. In principle, one
can add second-order terms to the constitutive relations and repeat the
derivation. Even without doing so, it is natural to expect that
higher-order terms will be ``generated'' by the hydrodynamic loop
corrections.

\item
We have assumed a particular convention for an off-equilibrium
definition of hydrodynamic variables, the Landau-Lifshitz ``frame''. The
correlation functions of the energy-momentum tensor and the current are
independent of our choosing one or another frame, and it is desirable to
have an effective action formulation where frame-invariance is
manifest.

\item
We have checked that, for smooth fields, the discrete-space expression
reproduces the continuum expressions we have written up to terms with 2
extra derivatives.  But the loop expansion under this effective action
will likely encounter UV divergences for which the discretization will
matter.  It is not clear to us how to deal with UV divergences in the
effective theory.  Also, the discretization we propose does not manifestly
preserve the symmetry of $T^{\mu\nu} = T^{\nu\mu}$.  It is not clear to us
whether this could cause any problems.

\item
The ghost part of the action has unusual properties; in particular, the
$F^2$ terms in \Eq{eq:Leff-1} are quartic in the ghosts, that is, there
are nonlinear ghost interactions.  The ghost part of the action could
presumably be made quadratic by introducing more auxiliary fields,
similar to what is done in other interacting fermion models.

\item
We have neglected the coupling of the hydrodynamic degrees of freedom to
external sources. While it is straightforward to couple the
action~\Eq{eq:finalaction} to the external gauge field and the
metric, one would like to have an effective action which provides us
with the full set of $n$-point real-time correlation functions. This
means that the action needs to be coupled to {\em two} sets of external
sources, corresponding to the two branches of the Schwinger-Keldysh
contour.
\end{itemize}
We plan to return to the above points in the future.

\subsection*{Acknowledgments}
The work of GM and PK was supported in part by NSERC of Canada. PR was
supported in part by the Sloan award No.\ BR2012-038 and by the DoE
Award No DE-SC0008027. We thank S.~de Alwis, P.~Arnold, K.~Jensen,
A.~Ritz, and T.~Schaefer for helpful discussions.

\bibliographystyle{utphys}
\bibliography{Seff}

\providecommand{\href}[2]{#2}\begingroup\raggedright\begin{thebibliography}{10}

\bibitem{LL6}
L.~D. Landau and E.~M. Lifshitz, {\em Fluid Mechanics}.
\newblock Pergamon, 1987.

\bibitem{Bhattacharyya:2008jc}
S.~Bhattacharyya, V.~E. Hubeny, S.~Minwalla, and M.~Rangamani, ``{Nonlinear
  Fluid Dynamics from Gravity},''
  \href{http://dx.doi.org/10.1088/1126-6708/2008/02/045}{{\em JHEP} {\bfseries
  02} (2008) 045}, \href{http://arxiv.org/abs/0712.2456}{{\ttfamily
  arXiv:0712.2456 [hep-th]}}.

\bibitem{Baier:2007ix}
R.~Baier, P.~Romatschke, D.~T. Son, A.~O. Starinets, and M.~A. Stephanov,
  ``{Relativistic viscous hydrodynamics, conformal invariance, and
  holography},'' \href{http://dx.doi.org/10.1088/1126-6708/2008/04/100}{{\em
  JHEP} {\bfseries 0804} (2008) 100},
  \href{http://arxiv.org/abs/0712.2451}{{\ttfamily arXiv:0712.2451 [hep-th]}}.

\bibitem{Kovtun:2012rj}
P.~Kovtun, ``{Lectures on hydrodynamic fluctuations in relativistic
  theories},'' \href{http://dx.doi.org/10.1088/1751-8113/45/47/473001}{{\em
  J.Phys.} {\bfseries A45} (2012) 473001},
\href{http://arxiv.org/abs/1205.5040}{{\ttfamily arXiv:1205.5040 [hep-th]}}.
%%CITATION = ARXIV:1205.5040;%%.

\bibitem{Pomeau197563}
Y.~Pomeau and P.~R\'esibois, ``Time dependent correlation functions and
  mode-mode coupling theories,''
  \href{http://dx.doi.org/10.1016/0370-1573(75)90019-8}{{\em Physics Reports}
  {\bfseries 19} no.~2, (1975) 63 -- 139}.

\bibitem{LL-1957}
L.~D. Landau and E.~M. Lifshitz, ``Hydrodynamic fluctuations,'' {\em JETP}
  {\bfseries 32} (1957) 618. ({\it Sov. Phys. JETP} {\bf 5}, 512 (1957)).

\bibitem{Forster:1977zz}
D.~Forster, D.~R. Nelson, and M.~J. Stephen, ``{Large-distance and long-time
  properties of a randomly stirred fluid},''
  \href{http://dx.doi.org/10.1103/PhysRevA.16.732}{{\em Phys.Rev.} {\bfseries
  A16} (1977) 732--749}.

\bibitem{PhysRevB.18.353}
C.~De~Dominicis and L.~Peliti, ``Field-theory renormalization and critical
  dynamics above ${T}_{c}$: Helium, antiferromagnets, and liquid-gas systems,''
  \href{http://dx.doi.org/10.1103/PhysRevB.18.353}{{\em Phys. Rev. B}
  {\bfseries 18} (1978) 353}.

\bibitem{Schutz:1970my}
B.~F. Schutz, ``{Perfect Fluids in General Relativity: Velocity Potentials and
  a Variational Principle},''
\href{http://dx.doi.org/10.1103/PhysRevD.2.2762}{{\em Phys.Rev.} {\bfseries D2}
  (1970) 2762--2773}.
%%CITATION = PHRVA,D2,2762;%%.

\bibitem{Brown:1992kc}
J.~D. Brown, ``{Action functionals for relativistic perfect fluids},''
  \href{http://dx.doi.org/10.1088/0264-9381/10/8/017}{{\em Class.Quant.Grav.}
  {\bfseries 10} (1993) 1579--1606},
\href{http://arxiv.org/abs/gr-qc/9304026}{{\ttfamily arXiv:gr-qc/9304026
  [gr-qc]}}.
%%CITATION = GR-QC/9304026;%%.

\bibitem{Carter:1994rv}
B.~Carter, ``{Axionic vorticity variational formulation for relativistic
  perfect fluids},''
\href{http://dx.doi.org/10.1088/0264-9381/11/8/009}{{\em Class.Quant.Grav.}
  {\bfseries 11} (1994) 2013--2030}.
%%CITATION = CQGRD,11,2013;%%.

\bibitem{Dubovsky:2011sj}
S.~Dubovsky, L.~Hui, A.~Nicolis, and D.~T. Son, ``{Effective field theory for
  hydrodynamics: thermodynamics, and the derivative expansion},''
  \href{http://dx.doi.org/10.1103/PhysRevD.85.085029}{{\em Phys.Rev.}
  {\bfseries D85} (2012) 085029},
\href{http://arxiv.org/abs/1107.0731}{{\ttfamily arXiv:1107.0731 [hep-th]}}.
%%CITATION = ARXIV:1107.0731;%%.

\bibitem{Bhattacharya:2012zx}
J.~Bhattacharya, S.~Bhattacharyya, and M.~Rangamani, ``{Non-dissipative
  hydrodynamics: Effective actions versus entropy current},''
  \href{http://dx.doi.org/10.1007/JHEP02(2013)153}{{\em JHEP} {\bfseries 1302}
  (2013) 153},
\href{http://arxiv.org/abs/1211.1020}{{\ttfamily arXiv:1211.1020 [hep-th]}}.
%%CITATION = ARXIV:1211.1020;%%.

\bibitem{Banerjee:2012iz}
N.~Banerjee, J.~Bhattacharya, S.~Bhattacharyya, S.~Jain, S.~Minwalla, {\em
  et~al.}, ``{Constraints on Fluid Dynamics from Equilibrium Partition
  Functions},'' \href{http://dx.doi.org/10.1007/JHEP09(2012)046}{{\em JHEP}
  {\bfseries 1209} (2012) 046},
\href{http://arxiv.org/abs/1203.3544}{{\ttfamily arXiv:1203.3544 [hep-th]}}.
%%CITATION = ARXIV:1203.3544;%%.

\bibitem{Jensen:2012jh}
K.~Jensen, M.~Kaminski, P.~Kovtun, R.~Meyer, A.~Ritz, {\em et~al.}, ``{Towards
  hydrodynamics without an entropy current},''
  \href{http://dx.doi.org/10.1103/PhysRevLett.109.101601}{{\em Phys.Rev.Lett.}
  {\bfseries 109} (2012) 101601},
\href{http://arxiv.org/abs/1203.3556}{{\ttfamily arXiv:1203.3556 [hep-th]}}.
%%CITATION = ARXIV:1203.3556;%%.

\bibitem{Andersson:2006nr}
N.~Andersson and G.~Comer, ``{Relativistic fluid dynamics: Physics for many
  different scales},'' \href{http://dx.doi.org/10.12942/lrr-2007-1}{{\em Living
  Rev.Rel.} {\bfseries 10} (2007) 1},
\href{http://arxiv.org/abs/gr-qc/0605010}{{\ttfamily arXiv:gr-qc/0605010
  [gr-qc]}}.
%%CITATION = GR-QC/0605010;%%.

\bibitem{Endlich:2012vt}
S.~Endlich, A.~Nicolis, R.~A. Porto, and J.~Wang, ``{Dissipation in the
  effective field theory for hydrodynamics: First order effects},''
  \href{http://dx.doi.org/10.1103/PhysRevD.88.105001}{{\em Phys.Rev.}
  {\bfseries D88} (2013) 105001},
\href{http://arxiv.org/abs/1211.6461}{{\ttfamily arXiv:1211.6461 [hep-th]}}.
%%CITATION = ARXIV:1211.6461;%%.

\bibitem{Grozdanov:2013dba}
S.~Grozdanov and J.~Polonyi, ``{Viscosity and dissipative hydrodynamics from
  effective field theory},''
\href{http://arxiv.org/abs/1305.3670}{{\ttfamily arXiv:1305.3670 [hep-th]}}.
%%CITATION = ARXIV:1305.3670;%%.

\bibitem{Kovtun:2003vj}
P.~Kovtun and L.~G. Yaffe, ``{Hydrodynamic fluctuations, long time tails, and
  supersymmetry},'' \href{http://dx.doi.org/10.1103/PhysRevD.68.025007}{{\em
  Phys.Rev.} {\bfseries D68} (2003) 025007},
\href{http://arxiv.org/abs/hep-th/0303010}{{\ttfamily arXiv:hep-th/0303010
  [hep-th]}}.
%%CITATION = HEP-TH/0303010;%%.

\bibitem{PeraltaRamos:2011es}
J.~Peralta-Ramos and E.~Calzetta, ``{Shear viscosity from thermal fluctuations
  in relativistic conformal fluid dynamics},''
  \href{http://dx.doi.org/10.1007/JHEP02(2012)085}{{\em JHEP} {\bfseries 1202}
  (2012) 085},
\href{http://arxiv.org/abs/1109.3833}{{\ttfamily arXiv:1109.3833 [hep-ph]}}.
%%CITATION = ARXIV:1109.3833;%%.

\bibitem{Kovtun:2011np}
P.~Kovtun, G.~D. Moore, and P.~Romatschke, ``{The stickiness of sound: An
  absolute lower limit on viscosity and the breakdown of second order
  relativistic hydrodynamics},''
  \href{http://dx.doi.org/10.1103/PhysRevD.84.025006}{{\em Phys.Rev.}
  {\bfseries D84} (2011) 025006},
  \href{http://arxiv.org/abs/1104.1586}{{\ttfamily arXiv:1104.1586 [hep-ph]}}.

\bibitem{Kapusta:2011gt}
J.~Kapusta, B.~Muller, and M.~Stephanov, ``{Relativistic Theory of Hydrodynamic
  Fluctuations with Applications to Heavy Ion Collisions},''
  \href{http://dx.doi.org/10.1103/PhysRevC.85.054906}{{\em Phys.Rev.}
  {\bfseries C85} (2012) 054906},
\href{http://arxiv.org/abs/1112.6405}{{\ttfamily arXiv:1112.6405 [nucl-th]}}.
%%CITATION = ARXIV:1112.6405;%%.

\bibitem{Kumar:2013twa}
A.~Kumar, J.~R. Bhatt, and A.~P. Mishra, ``{Fluctuations in Relativistic Causal
  Hydrodynamics},''
\href{http://arxiv.org/abs/1304.1873}{{\ttfamily arXiv:1304.1873 [hep-ph]}}.
%%CITATION = ARXIV:1304.1873;%%.

\bibitem{Murase:2013tma}
K.~Murase and T.~Hirano, ``{Relativistic fluctuating hydrodynamics with memory
  functions and colored noises},''
\href{http://arxiv.org/abs/1304.3243}{{\ttfamily arXiv:1304.3243 [nucl-th]}}.
%%CITATION = ARXIV:1304.3243;%%.

\bibitem{Young:2013fka}
C.~Young, ``{Numerical integration of thermal noise in relativistic
  hydrodynamics},''
\href{http://arxiv.org/abs/1306.0472}{{\ttfamily arXiv:1306.0472 [nucl-th]}}.
%%CITATION = ARXIV:1306.0472;%%.

\bibitem{LL9}
E.~M. Lifshitz and L.~P. Pitaevskii, {\em Statistical Physics, Part 2}.
\newblock Pergamon, 1980.

\bibitem{Arnold:1999va}
P.~B. Arnold, ``{Symmetric path integrals for stochastic equations with
  multiplicative noise},''
  \href{http://dx.doi.org/10.1103/PhysRevE.61.6099}{{\em Phys.Rev.} {\bfseries
  E61} (2000) 6099--6102},
\href{http://arxiv.org/abs/hep-ph/9912209}{{\ttfamily arXiv:hep-ph/9912209
  [hep-ph]}}.
%%CITATION = HEP-PH/9912209;%%.

\bibitem{Zanella:2002hh}
J.~Zanella and E.~Calzetta, ``{Renormalization group and nonequilibrium action
  in stochastic field theory},''
  \href{http://dx.doi.org/10.1103/PhysRevE.66.036134}{{\em Phys.Rev.}
  {\bfseries E66} (2002) 036134},
\href{http://arxiv.org/abs/cond-mat/0203566}{{\ttfamily arXiv:cond-mat/0203566
  [cond-mat]}}.
%%CITATION = COND-MAT/0203566;%%.

\end{thebibliography}\endgroup

\end{document}